\documentclass[preprint,proceedings]{rmaa}

\SetYear{2004}
\SetConfTitle{The Environment and Evolution of Binary and Multiple Stars}

\title{On the Formation of Brown Dwarfs}

\author{Ing-Guey Jiang\altaffilmark{1}, G. Laughlin\altaffilmark{2}
  and D.N.C. Lin\altaffilmark{2}}

\altaffiltext{1}{National Central Univ., Taiwan. jiang@astro.ncu.edu.tw}
\altaffiltext{2}{University of California, USA. laughlin, lin@ucolick.org}

\shortauthor{Jiang, Laughlin and Lin}
\shorttitle{Brown Dwarfs}
\fulladdresses{
\item Ing-Guey Jiang: Institute of Astronomy, National Central University,
Taiwan (\email{jiang@astro.ncu.edu.tw}).
\item G. Laughlin and D.N.C. Lin: UCO/Lick Observatory, University of 
 California, Santa Cruz, CA 95064, USA (\email{laughlin, lin@ucolick.org}).}
\listofauthors{Ing-Guey Jiang, G. Laughlin, D.N.C. Lin}
\indexauthor{Jiang, I.-G.}
\indexauthor{Laughlin, G.}
\indexauthor{Lin, D.N.C.}

\addkeyword{Stars - brown dwarfs}

\abstract{The observational properties of brown dwarfs
pose challenges to the theory of star formation.  Because their mass
is much smaller than the typical Jeans mass of interstellar clouds,
brown dwarfs are most likely formed through secondary fragmentation
processes, rather than through the direct collapse of a molecular
cloud core. In order to prevent substantial post-formation mass
accretion, young brown dwarfs must leave the high density formation
regions in which they form.  We propose here that brown dwarfs are
formed in the optically thin outer regions of circumbinary disks.
Through post-formation dynamical interaction with their host binary
stars, young brown dwarfs are either scattered to large distance or
removed, with modest speed, from their cradles.}

\resumen{Las propiedades de las enanas caf\'es presentan problemas para la teor\'\i a de
la formaci\'on estelar.  Como sus masas son mucho menores que la masa de Jeans
de las nubes interestelares, las enanas caf\'es se forman probablemente por
fragmentaci\'on secundaria, y no por el colapso directo del n\'ucleo de la nube
molecular.  Para evitar la acreci\'on de masa posterior, las j\'ovenes enanas
caf\'es deben salir de las regiones de alta densidad donde se formaron.
Proponemos que las enanas caf\'es se forman en las regiones externas, \'opticamente
delgadas, de los discos circunbinarios.  Las interacciones din\'amicas
subsecuentes con sus binarias causan que las enanas caf\'es sean dispersadas a grandes
distancias, o que escapen del sistema a baja velocidad.}

\begin{document}
\maketitle 

\section{Introduction}
Brown dwarfs are entities with mass below that require for hydrogen
burning to ignite ($< 0.075 M_\odot$) and above that associated with
gaseous giant planets ($\sim$ a few Jupiter mass, $M_J$). 
Although the existence of brown dwarfs was proposed by Kumar in 1963, 
their cool dim nature consigned them to a strictly theoretical status for 
more than three decades.  Recently, however, improved observational 
capabilities have led to the discovery of many brown dwarfs, prompting a 
renaissance in our understanding of these objects.
%These searches for brown dwarfs reveal that (a) they are as numerous
%as M dwarfs both in the field and in young stellar clusters, (b) they
%are occasionally wide companions but rarely close companions to main
%sequence stars, and (c) brown dwarf-brown dwarf pairs tend to be close
%binary systems with separations less than a few AU.
Because the mass of 
brown dwarfs is much smaller than the usual Jeans mass for a typical 
molecular cloud, Lin et al. (1998) claimed that the encounter between two 
protostellar discs might increase the Jeans mass locally. Reipurth \& Clarke 
(2001) suggested that brown dwarfs are substellar objects because they have 
been ejected from newborn multiple systems. They use a simple model of 
timescales to show that this could happen. Bate et al. (2002) used a smoothed 
particle hydrodynamics code to show that brown dwarfs could be formed through 
the collapse and fragmentation of a turbulent molecular cloud and thus 
confirmed what was suggested by Reipurth \& Clarke (2001).

However, if all brown dwarfs have to be ejected to avoid accreting 
too much mass, one might ask how it would be possible that they are 
wide companions of solar-type stars sometimes. In this paper,
we propose a formation scenario for brown dwarfs which provides a
natural explanation for the current observational situation.
We suggest that brown dwarfs could be formed through disc fragmentation
and we study the ``escape zone'' where the brown dwarfs could be ejected
and become a field star.

In principle, we demonstrate that in some cases, the brown dwarfs get ejected
but in other cases, they could become long-period companion. 
We also study the criteria that the brown dwarf pair can survive.
However, we do not intend to study the problem of brown dwarf desert here.

\section{Stability of brown dwarf companion around binary stars}

We now investigate the orbital stability and evolution of brown dwarfs
formed in circumbinary rings.  These companions are perturbed by the
tidal disturbance of the binary star's gravitational potential.
For computational convenience, we assume
these newly fragmented brown dwarfs quickly become centrally condensed
and the residual gas does not contribute significantly to the
gravitational potential such that the dynamics of the system may be
described by a few-body approximation.

In this section, we further simplify the interaction procedure to a
three-body (the host binary star plus a brown dwarf) problem.  The
mass of each brown-dwarf fragment is sufficiently small that they do
not significantly perturb each other on the short term of several thousand 
binary periods. In order to make a direct comparison with some existing
results, we first treat the brown dwarfs as massless particles. 
But, in general, we carry out full 3-body integration in which 
contribution due to the finite mass of the brown dwarf is included.

We consider a range of ratio ($\mu = M_1 / (M_1 + M_2)$) of masses
($M_1$ and $M_2$) for the two components of the binary stars.
Following the approach by Holman \& Wiegert (1999), we consider, for
the host binary system, a range of orbital eccentricity ($e_\ast$).  
The semi-major axis of the binary is set to be unity such that all other 
length scales are scaled with its physical value. We also adopt 
$G (M_1 + M_2) =1$ such that the binary orbital period is $2 \pi$.

Since we assume they are formed in circumbinary rings, we consider brown 
dwarfs with orbital semi major axis larger than that of the binary system.  
At the onset of the computation, all three stars are located at their apocenter
with respect to their common center of mass. It is possible that the fragments 
may have a range of orbital eccentricity ($e_b$). Here, we consider two 
limiting eccentricity for the brown dwarf ($e_b = 0$ and 0.4).  We also 
set the ratio ($\mu_b$) of brown dwarf's mass to that of the binary to be 
0.05 and we assume brown dwarfs rotate in the same direction as the orbit 
of the binary secondary.

\subsection{Ejection criteria}
We choose a range of initial orbital semi major axis for the brown dwarf for 
the $e_b = 0$ case.  For each set of model parameters, we adopt four values 
for the brown dwarf's angle of apocenter, $0^{\circ}$, $90^{\circ}$, 
$180^{\circ}$ and $270^{\circ}$ with respect to that of the binary system.

In all our models, the orbital semi-major axes of the brown dwarf is
larger than that of the binary system.  We are primarily seeking a
critical initial semi-major axis ($a_c$), larger than which the brown
dwarf survives the binary system's perturbation within a timescale
$T_d$.  Our definition of survival is that the distance from the center of
mass of the system to the brown dwarf (starting with all four values of
the apocentric arguments) must be smaller than a critical value $R_d$.  
The value of $R_d$ is arbitrarily set to be 25 binary separations.  In 
order to compare with the results of Holman \& Wiegert (1999), we choose 
$T_d = 10^4$ binary period. Based on several test runs, we find that the 
value of $a_c$ does not change significantly if $T_d$ is increased to 
$10^6$ binary period. Thus, we find $a_c$ to be a useful parameter to 
classify our results (Dvorak 2004). Although we use a totally different 
numerical scheme as Holman \& Wiegert (1999), we are able to precisely 
reproduce the results in Table 7 of their paper when we set the mass of 
brown dwarf to be zero as they have. But in general, we choose $\mu_b = 0.05$.

\begin{figure}[!t]
  \includegraphics[width=.9\columnwidth]{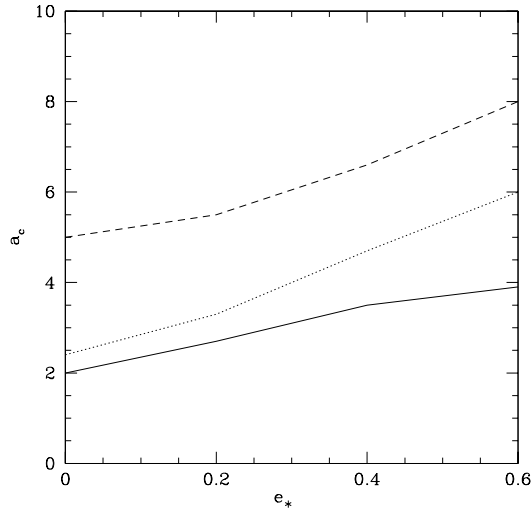}
  \caption{  The critical semi major axis as function of eccentricity
when the mass ratio of the binary system $\mu=0.1$. The solid line represents 
the result for models with brown dwarfs been treated as massless test 
particles which initially move on circular orbits. These models are 
analogous to those obtained by  Holman \& Wiegert (1999). The dotted
line represents the results for models in which  brown dwarfs are assigned 
with finite mass $\mu_b= 0.05$. The brown dwarfs are assumed to have  
circular orbits about the center of mass of the binary system initially.  
The dash line represents the results of models in which brown dwarfs have 
mass $\mu_b=0.05$ and have initial eccentricity $e_b = 0.4$}
  \label{Figure}
\end{figure}

\begin{figure}[!t]
  \includegraphics[width=.9\columnwidth]{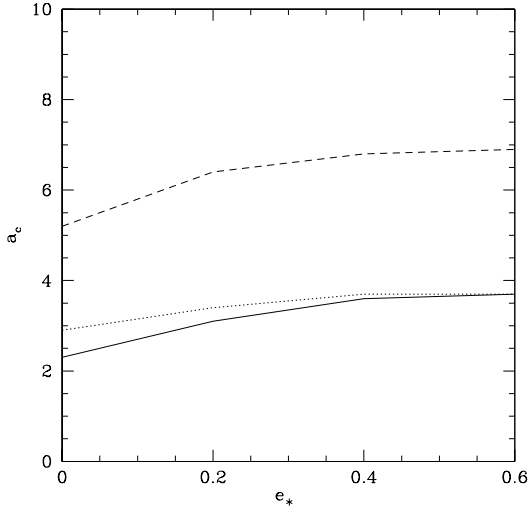}
  \caption{The same as in Fig.1 except the mass ratio of the 
binary system $\mu = 0.5$.}
  \label{Figure}
\end{figure}

From these models, we find that the brown dwarf's ``escape zone'' (with 
semi-major axis $a < a_c$) is expanded slightly when they have finite mass 
(see Fig.1 \& 2 for the comparison.)  The expansion of the ``escape zone'' 
is larger for $\mu=0.1$ cases than that for the $\mu=0.5$ cases because 
the motion of secondary star is more affected by the finite mass of the 
brown dwarf. This effect is particularly noticeable for the $\mu=0.1$ and 
$e_\ast=0.6$ case where $a_c$ is expanded from $3.9$ of massless particles 
to $6.0$ of brown-dwarfs with $\mu_b =0.05$. The hydrodynamical simulations 
indicate that fragmentation of circumbinary disks occurs primarily at 
around 1-2 binary separation away from their center of mass because this 
is the location where disk gas may accumulate as a consequence of binary 
star's tidal torque. Thus, {\it most of the low-mass fragments formed in 
the circumbinary rings have a high probability of being ejected by the 
gravitational perturbation of their host binary systems.}

We now consider a series of models with $e_b = 0.4$ while all other parameters
are similar to those for the $e_b = 0$ case. The increases in $a_c$ in Fig. 1 
and Fig. 2 clearly indicate that brown dwarfs with eccentric orbits are
definitely less stable than those with circular orbits.

\subsection{Large radial excursion of marginally stable systems.}

In general, the binary systems and the fragments formed in unstable
circumbinary rings have non circular orbits. Thus, most
brown dwarfs formed close to the binary are likely to be 
ejected.  But, brown dwarfs with initial semi major axis $a_b \sim a_c$,
can be scattered to large distances from the center of mass of the
system without escaping from its gravitational potential.  

We illustrate three such examples each with $a_b \sim a_c$.  In model 1, 
we choose $\mu = 0.1$, $e_\ast = 0.4$, $a_b = 4.4$, $e_b =0$ and the argument 
of brown dwarf's apogee, $\theta_b = 90^{\circ}$.  In this case, the brown 
dwarf reaches to 200 binary separation by the end of simulation, i.e. 
$t = T_d$. In model 2, ($\mu = 0.1$, $e_\ast = 0.6$, $a_b = 7.8$, $e_b = 0.4$,
and $\theta_b = 180^{\circ}$), the brown dwarf's orbit expands to 50 times the 
binary's initial separation at $t \sim 0.6 T_d$. But subsequently at $t = T_d$,
the extent of the brown dwarf's radial excursion is reduced to approximately 
its initial value.  In model 3, ($\mu = 0.5$, $e_\ast = 0.2$, $a_b = 6.3$, 
$e_b = 0.4$, and $\theta_b = 270^{\circ}$), the excursion reaches 100 initial 
binary separations.  These examples indicate that wide and marginally stable 
orbits exist and that {\it under some marginal circumstances, brown dwarfs 
can be scattered to large distance from but remain bound to some
main-sequence binary stars.} The recent discovery of a brown dwarf 
candidate at a distance of $\sim 100$ AU from a young binary star, TWA 6 
Hyd may be examples of such a system.

\subsection{Ejection speed of escapers} 
Brown dwarfs with $a_b < a_c$ are ejected from the gravitational potential of 
their host binary system.  We now examine their escape speed by series 
of calculations. 

From the distribution of the escape speed, we found  that the ejection speed 
is typically half the orbital speed of the binary. 
For $R_b \sim 3 \times 
10^{15}$ cm, $a \sim R_b$, 
and the total mass of binary system $\sim 1 M_\odot$,
the binary's orbital speed would need to be $\sim 3-5$ km s$^{-1}$. Our results
show that the escaping speed of the brown dwarf ejecta is $\sim 1-3$ 
km s$^{-1}$. In a young stellar cluster, such as the Orion, this ejection 
speed is a fraction of the velocity dispersion of the cluster which is in a
dynamical equilibrium.  Thus, {\it brown dwarfs ejected from the close
proximity of the binary would not generally escape the gravitational
potential of the cluster.}  This result is consistent with the large
concentration of brown dwarfs in young stellar clusters such as the
Orion complex (Lucas \& Roche 2000).

\section{Formation of brown dwarf pairs}
Indeed, close brown dwarf binaries have been found (Koerner et al. 1999), 
but these systems are generally not orbiting around some other binary main 
sequence stars. Similar to single brown dwarfs, close binary brown dwarfs 
may also be strongly perturbed by the gravity of the binary and be ejected.

\subsection{Survival of pairs}
In order to test the survival probability of the brown dwarf binaries, we 
first place a massless test particle to simulate the dynamics of a secondary 
companion around the brown dwarf (For some interesting cases, a series of 
models with masses for both brown dwarf companions are included). This 
approximation allows us to first explore the range of parameters which may be 
favorable for the survival of the brown dwarf pairs.  Based on the results in 
Fig.1 and Fig. 2, we can identify the range of model parameters which leads to 
ejection of brown dwarf fragments.  As a test, we adopt $\mu = 0.5$ for the 
binary star and choose $\mu_b = 0.05$, $a_b = 2.3$ and apogee at $0^{\circ}$ 
for the primary of the brown dwarf binary.  The secondary of the brown dwarf 
binary is assigned an initial semi major axis $2.5$ and apogee at $0^{\circ}$.
Thus the separation of the brown-dwarf pair is $0.2$ which is inside the Roche
radius of the primary ($R_R = (\mu_b/3)^{1/3} = 0.25$).  We also assume that 
the center of mass of the brown-dwarf pair is initially on a circular orbit 
around the center of mass of the binary system and the brown-dwarf secondary 
is on a circular orbit around the brown-dwarf primary.

Then we try four cases of different eccentricities of central binary stars 
relative to each other: $e_\ast = 0.0,~0.2,~0.4,~0.6$. Our results
indicate that the brown-dwarf pairs would survive their ejection from
the neighborhood of the binary in the low-eccentricity 
($e_\ast = 0.0$ and $e_\ast = 0.2$) limit.  But, in the limit that the 
binary system has a large eccentricity ({\it i.e.} for the
$e_\ast = 0.4$ and $e_\ast = 0.6$ models), the brown dwarf pairs have
a tendency to become dissociated. Therefore it is plausible that brown 
dwarf pairs may remain bound to each other during their ejection from the 
binary system's gravitational potential. But it is also possible for 
their ejection to produce two freely floating single brown dwarfs.

The above approximation provides a useful tool for us to identify the
range of parameters which allows a brown dwarf pairs to survive.  The
massless approximation for the secondary is applicable to brown dwarf
binaries with extreme mass ratios.  We now takes the next iteration by
replacing the ($\mu_b = 0.05$) primary and a massless secondary
brown-dwarf pair with a system of two equal-mass ($\mu_b = 0.025$)
brown dwarf companions.  We find, with identical four sets of model
parameters as above, brown dwarf binaries remain intact as they are
ejected by their host binary stars.

We also enlarge the separation of the brown-dwarf binary to 0.3, which
is larger than its Roche radius.  Again all four sets of initial
conditions are used.  In all these cases, the brown dwarf binary is
always disrupted during its close encounters with the binary star.

\subsection{Pair capture}

We now explore the possibility that both brown 
\adjustfinalcols
dwarfs were formed as 
single objects and that they 
%\adjustfinalcols
 have captured each other to become a binary.
In order to evaluate this probability, we repeat the earlier 
 simulation
in which 24 particles are placed in a ring around the binary system.
The main difference from the earlier models is that a mass of $\mu_b = 0.05$ 
is assigned to each particles.  The corresponding Roche radius for each 
individual particle is $\sim 0.25 $ which is comparable to their initial 
separation. All the particles are ejected from the gravitational potential 
of the binary system but no particle captured any other particle. We did not 
see any case in which the capture happened. We thus conclude that this 
second scenario is very unlikely.

\section{Summary}

For single brown dwarf satellites around binary systems, our dynamical 
calculations show that when they are formed in dynamically unstable 
regions, they are likely to be ejected from the gravitational potential of 
the binary system. These results provide an explanation for the common 
sighting of field brown dwarfs.

For binary brown dwarf satellite pairs, the calculations of four body (main
sequence binary with brown dwarf pair) interaction show that these 
brown-dwarf pairs can remain to be bound to each other during the 
ejection if their initial separation is well within their Roche radius
(which is typically $R_R \sim 0.2-0.25$ binary separation).

\acknowledgements We are grateful for friendly conversation with Matthew Bate,
Michael Sterzik and Bo Reipurth. Particularly, Michael Sterzik and Bo Reipurth
pointed out that the words ``Main-Sequence'' probably should be replaced by 
``Solar-Type Stars'' in the viewgraphs of the oral presentation of this paper.


\begin{thebibliography}

\bibitem{B02} Bate, M. R., Bonnell, I. A., Bromm, V. 2002, MNRAS, 332, L65

\bibitem{D} Dvorak, R. 2004, this volume

\bibitem{H02} Holman, M. J. \& Wiegert, P. A. 1999, \aj, 117, 621  

\bibitem{K01} Koerner, D. W., Kirkpatrick, J. D., McElwain, M. W. 
 \& Bonaventura, N. R. 1999, \apj, 526, L25 

\bibitem{L01} Lin, D. N. C., Laughlin, G., Bodenheimer, P., Rozyczka, M. 1998,
 Science, 281, 2025

\bibitem{L02} Lucas, P. W., Roche, P. F. 2000, MNRAS, 314, 858 

\bibitem{R02} Reipurth, B., Clarke, C. J. 2001, AJ, 122, 432 

\end{thebibliography}
\end{document}